\def \deg {\ensuremath{^\circ} }

\documentclass[aps,prl,preprint,superscriptaddress]{revtex4}

\usepackage{graphicx}
\usepackage{dcolumn}
\usepackage{bm}
\usepackage[english]{babel}

\begin{document}

\preprint{V1 -- \today{}}

\title{Triggering and probing of phase-coherent spin packets by time-resolved spin transport across an Fe/GaAs Schottky barrier}

\author{L. R. Schreiber}
\author{C. Schwark}
\author{S. Richter}
\author{C. Weier}
\author{G. G\"untherodt}
\affiliation{II. Institute of Physics, RWTH Aachen University, 52074
Aachen, Germany} \affiliation{JARA: Fundamentals of Future
Information Technology, 52074 Aachen, Germany}
\author{C. Adelmann}
\affiliation{Department of Chemical Engineering and Materials
Science, University of Minnesota, Minneapolis, Minneapolis 55455}
\author{C. J. Palmstr{\o}m}
\affiliation{Department of Electrical and Computer Engineering and Department of Materials, University of California at
Santa Barbara, Santa Barbara, California 93106}
\author{X. Lou}
\author{P. A. Crowell}
\affiliation{School of Physics and Astronomy, University of
Minnesota, Minneapolis, Minnesota 55455}
\author {B. Beschoten}
\email[]{bernd.beschoten@physik.rwth-aachen.de} \affiliation{II.
Institute of Physics, RWTH Aachen University, 52074 Aachen, Germany}
\affiliation{JARA: Fundamentals of Future Information Technology,
52074 Aachen, Germany}

\date{\today}

\begin{abstract}

Time-resolved electrical spin transport is used to generate and
probe spin currents in GaAs electrically. We use high bandwidth current
pulses to inject phase-coherent spin packets
from Fe into $n$-GaAs. By means of time-resolved
Faraday rotation we demonstrate that spins are injected with a clearly defined phase by the observation
of multiple Larmor precession cycles. We furthermore show
that spin precession of optically created spin packets in $n$-GaAs
can be probed electrically by spin-polarized photo-current pulses.  The injection and detection
experiments are not direct reciprocals of each other.  In particular, we find that interfacial spin accumulation
generated by the photocurrent pulse plays a critical role in time-resolved electrical spin detection.
\end{abstract}

\pacs{72.25.Fe, 76.30.Pk, 42.50.Md, 78.47.-p}
\keywords{XXX}
\maketitle

Spintronic devices which are based on coherence of an electron
spin ensemble require a means to trigger all individual
spins with a well-defined phase. Both temporal and spatial phase triggering can easily be
achieved by optical orientation using either circularly
\cite{Kikkawa98, Kikkawa99, Kimel2001PRB, Kato03,
Greilich2006Science, Meier2007NaturePhys} or linearly polarized
\cite{Schmalbuch2010PRL} laser pulses. This triggering results in a
macro-phase of the spin ensemble. In a transverse magnetic field all
individual spins of this ensemble start to precess from the
identical spin orientation, i.e. with the same phase. The coherence
of the spin ensemble can be probed by spin precession using
time-resolved (TR) magneto-optical techniques \cite{Kikkawa98,
Kikkawa99, Kimel2001PRB, Kato03, Greilich2006Science,
Meier2007NaturePhys, Schmalbuch2010PRL}. In dc spin transport
experiments, however, there is only a spatial phase triggering of
the spin orientation after electrical spin injection, which is
defined by the magnetization direction of the ferromagnetic injector
\cite{Ohno99Nature, Zhu01, Hanbicki02, Jiang05, Adelmann05,
Kotissek2007, Ciorga2009}. After injection, individual spins start
to precess in a transverse magnetic field. This results in a rapid
depolarization of the steady-state spin polarization (the Hanle
effect), because spins are injected continuously in the time domain.
During precession the phase relation is partially preserved when
there is a well-defined transit time of a spin between the source
and detector \cite{Crooker05Science, Appelbaum07}. While several
spin precessions have been observed in spin transport devices based
on silicon \cite{Appelbaum07, Huang07, Li08}, only very few
precessions are seen in drift-based spin transport devices
\cite{Crooker05Science, Kato05APL, Lou07}. Despite recent progress
in realizing all-electrical spintronic devices, electrical phase
triggering and time-resolved electrical readout have not been achieved.

In this Letter we report on TR electrical spin injection (TRESI) and
electrical detection of spin currents across an Fe/GaAs Schottky
barrier by combining ultrafast electrical and optical techniques. Our
device consists of a highly doped Schottky tunnel barrier formed
between an epitaxial iron (Fe) and a (100) $n$-GaAs layer. We have
chosen this device design for three reasons: (I) the Schottky
barrier profile guarantees large spin injection efficiencies
\cite{Hanbicki02, Adelmann05}, (II) the $n$-GaAs layer is Si doped
with carrier densities near the metal-insulator transition ($n=2-4
\times 10^{16}$ cm$^{-3}$), which provides long spin dephasing times
$T_2^*$ \cite{Kikkawa98, Dzhioev02, Schmalbuch2010PRL} and (III) the
Fe injector has a two-fold magnetic in-plane anisotropy
\cite{Crooker05Science}, which allows for a non-collinear alignment
between the external magnetic field direction and the magnetization
direction of the Fe layer and thus the polarization direction of the
electrically injected spin packets. This non-collinear alignment
allows to induce Larmor precession of the spin ensemble.

Our measurement setup for TRESI and the sample geometry are depicted
in Fig.~1(a) (see also Ref. \cite{supplementary}). The magnetic easy
axis (EA) of the Fe layer is oriented along the GaAs [110] ($\pm x$
direction)(see Fig. 2(a) in Ref. \cite{supplementary}). The samples are
mounted in a magneto-optical cryostat with a magnetic field $B$
oriented along the $\pm z$ direction. We apply a current pulse train
using high frequency contacts. Linearly polarized ps laser pulses at
normal incidence to the sample plane and phase-locked to the
electrical pulses monitor the $\pm y$ component of spins injected in
the GaAs by detecting the Faraday rotation angle $\theta_F$. The
time-delay $\Delta t$ between the current pump pulses and optical
probe pulses is adjusted by an electronic phase shifter.

We first use static measurements of $\theta_F$ to demonstrate
electrical spin injection from Fe into $n$-GaAs [Fig.~1(b)]. At
reversed bias and $B_z=0$~T, injected spins are oriented parallel to
the EA of the Fe layer yielding $\theta_F=0$. At small $B_z$, spins
start to precess towards the $y$-direction yielding $\theta_F\neq
0$. Changing the sign of $B_z$ inverts the direction of the spin
precession which results in a sign reversal of $\theta_F$. As
expected \cite{Crooker05Science}, the direction of spin precession
also inverts when the magnetization $\mathbf{M}_{Fe}$ is reversed
(see red curve in Fig.~1(b)). $\theta_F$ approaches zero at large
$B_z$, since the continuously injected spins dephase due to Larmor
precession causing Hanle depolarization.

\begin{figure}[tbp]
\includegraphics{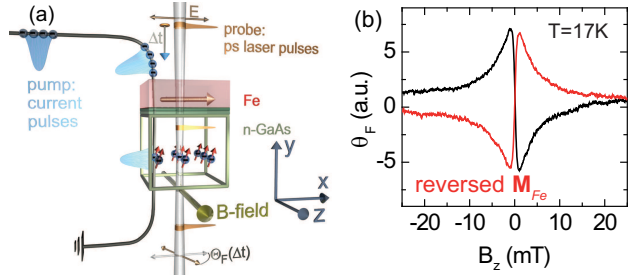}
\caption{ \label{fig1} (color online). (a) Schematic of the
time-resolved electrical spin injection experiment. (b) Faraday
rotation $\theta_F$ from dc electrical spin injection vs $B_z$.
Spins are probed in $n$-GaAs along $y$ before (black line) and after
(red line) reversing the Fe magnetization $\mathbf{M}_{Fe}$. }
\end{figure}

For TRESI, we now apply current pulses with a width of $\Delta w =
2$~ns and a repetition time of $T_{rep} = 125$~ns with
$T_{rep}>T_2^*$. The corresponding TRFR data are shown in Figs.
2(a), 2(b) at various $B_z$ and $T=17$~K. Most strikingly, we
observe precession of the injected spin packets
at the Larmor frequency $\omega_L$, demonstrating that the current pulses trigger spin currents with a well-defined
phase. It is apparent that the amplitude of $\theta_F$ diminishes with increasing $|B_z|$, and the oscillations in
$\theta_F$ are not symmetric about the zero base line (see black
lines in Fig. 2a as guides to the eye). For quantitative analysis we
use the ansatz
\begin{eqnarray}
 \nonumber \theta_F(\Delta t, B_z)&=& A(B_z) \exp\left(-\frac{\Delta t}{T_2^*(B_z)}\right) \sin(\omega_L \Delta
 t + \phi)+ \\
                &+& A_{bg}(B_z) \exp\left(-\frac{\Delta t}{\tau_{bg}}\right), \label{eq:simplebk}
\end{eqnarray}
\noindent where the second term accounts for the non-oscillatory
time dependent background. The fits to the data are shown in
Fig.~2(a) as red curves. We determine a field independent $\tau_{bg}
= 8 \pm 2$~ns and deduce $|g| = 0.42 \pm 0.02$ from $\omega_L$,
demonstrating that the spin precession is detected in the $n$-GaAs layer
\cite{Kikkawa98}. The extracted spin dephasing times $T_2^*(B_z)$
and amplitudes $A(B_z)$ are plotted in Figs.~2(c) and 2(d),
respectively. The longest $T_2^*(B_z)$ values, which exceed 65~ns,
are obtained at small $B_z$. The observed 1/$B_z$ dependence of
$T_2^*$ (see red line in Fig.~2(c)), which indicates inhomogeneous
dephasing of the spin packet, is consistent with results obtained
from all-optical TR experiments on bulk samples with similar doping
concentration \cite{Kikkawa98}. On the other hand, the strong
decrease of $A(B_z)$ (Fig. 2(d)) has not previously been observed.

\begin{figure}[tbp]
\includegraphics{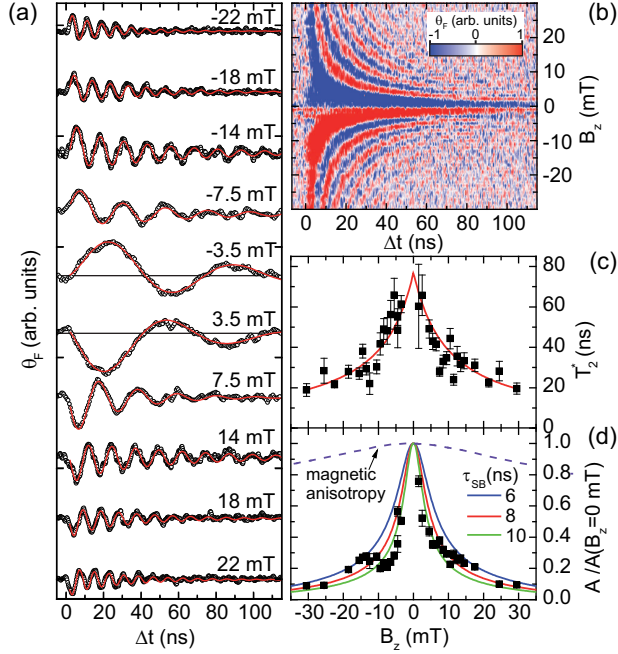}
\caption{ \label{fig2} (color online). (a) Faraday rotation
$\theta_F$ vs delay $\Delta t$ for various magnetic fields $B_z$ at
$T=17$~K with vertical offsets for clarity.  The red lines are fits to
the data. (b) False color plot of $\theta_F$ vs $\Delta t$ and
$B_z$. (c) Spin dephasing time $T_2^*(B_z)$ and (d) normalized
$\theta_F$ amplitudes $A$ vs $B_z$. The red solid lines in (c) are
least-squares fits to the data for $\Delta t> 2$~ns and solid lines
in (d) are simulations for different $\tau_{SB}$. The dashed line is
the expected decrease of $A$ due to the rotation of the magnetization of the
Fe layer.}
\end{figure}

The $A(B_z)$ dependence might be caused by $B_z$ acting on the
direction of the magnetization $\mathbf{M}_{Fe}$ of the Fe injector.
Increasing $B_z$ rotates $\mathbf{M}_{Fe}$ away from the easy
($x$-direction) towards the hard axis ($z$-direction). This rotation
diminishes the $x$-component of the injected spin packet, which
would result in a decrease of $A(B_z)$. We calculated this
dependence (see dashed line in Fig. 2(d)) for a macro-spin along
$\mathbf{M}_{Fe}$ using in-plane magnetometry data of the Fe
layer\cite{supplementary}. The resulting decrease, however, is by
far too small to explain our $A(B_z)$ dependence. To summarize,
there are two striking observations in our TRESI experiments: (I)
the strong decrease of $A$ vs $B_z$ and (II) the non-oscillatory
background $\tau_{bg}$ in $\theta_F$ independent of $B_z$. As both
have not been seen in all-optical time-resolved experiments, it is suggestive to
link these properties to the dynamics of the electrical spin
injection process.

\begin{figure}[tbp]
\includegraphics{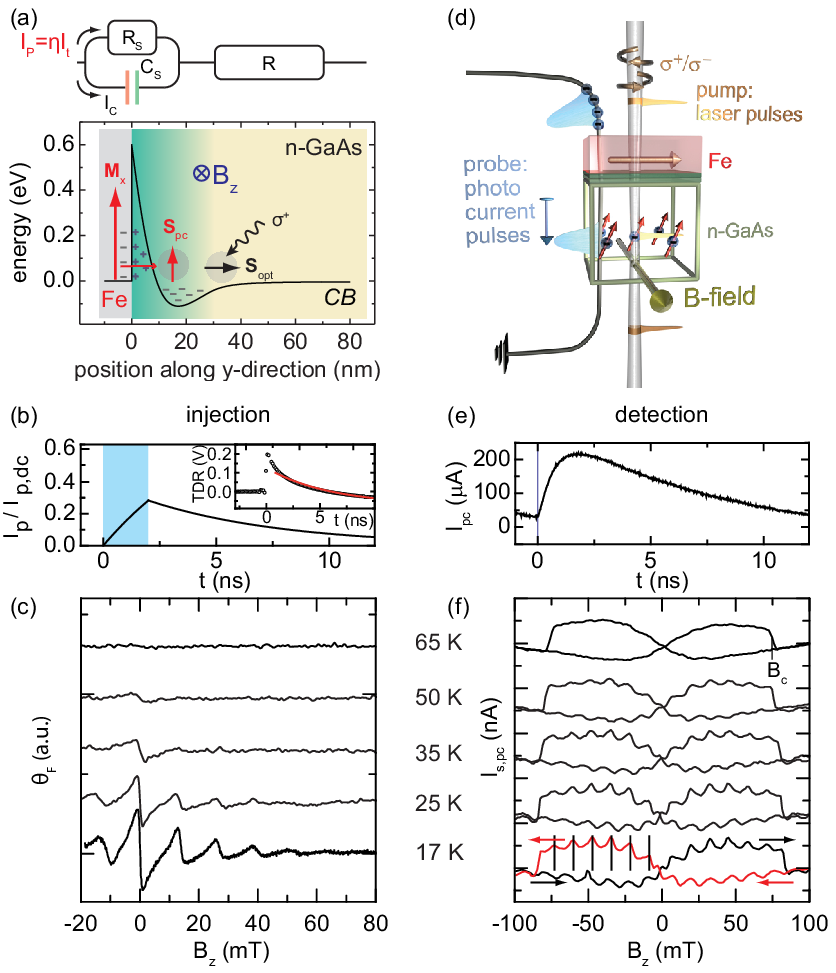}
\caption{ \label{fig3} (color online) (a) Simple equivalent network
of sample and energy diagram of conduction band. (b) Simulation of
the evolution of the spin-polarized tunnel current through the
Schottky barrier triggered by a 2 ns long current pulse (light
blue). (inset) TDR of voltage step function (-0.5~V) with
least-squares exponential fit (red solid line). (c) RSA measurements
of $\theta_F$ vs $B_z$ at various $T$ with vertical offsets for
clarity. (d) Schematic of optical pump/photo-current probe
experiment. (e) Measured photo-voltage pulse. (f) RSA measurements
of spin photo-current at various $T$.  }
\end{figure}

During TRESI, the depletion layer at the Schottky barrier (see band
diagram in Fig.~3(a)) acts like a capacitance. When a current pulse
is transmitted through the barrier, this capacitance will be charged
and subsequently discharged. The charging and discharging can
directly be measured by time-domain reflectometry (TDR) (see ref.
\cite{supplementary} for details) using a sampling oscilloscope. For
TDR we apply a voltage step to the sample with an amplitude of
-0.5~V and a rise time of 100~ps at $T=20$~K. The time-evolution of
the reflected voltage step is shown in the inset of Fig.~3(b). Note
that there is a significant temporal broadening of the voltage step
with a time constant of 6~ns. We obtain a similar time constant of
6~ns for the discharging process (not shown).

To further link these properties to the pulsed electrical spin
injection process, we depict a simple equivalent network of the
sample in Fig.~3(a). In the reverse-bias regime, the Schottky
contact can be modeled by a Schottky capacitance $C_s$ and a
parallel tunnel-resistance $R_s$. The underlying $n$-GaAs detection
layer is represented by a resistance $R$ in series. We assume the
displacement current $I_c$ to be unpolarized, while the tunneling
current $I_t$ carries the spin polarized electrons. The spin current
$I_p = \eta I_t$ is given by the spin injection efficiency $\eta$.
The charging and discharging of the Schottky capacitance is thus
directly mapped to the temporal evolution of the spin current. $I_p$
increases after the current pulse is turned on, whereas it decreases
after the pulse is turned off after time $\Delta w$, i.e. during the
discharge of $C_s$. If $C_s$, $R_s$ and $\eta$ are approximately
bias-independent, the increase and decrease of $I_p$ is a
simple exponential with a time constant $\tau_{SB}=C_S/
(\frac{1}{R}+\frac{1}{R_S})$ as illustrated in Fig. 3(b) for a pulse
width of $\Delta w=2$~ns and $\tau_{SB}=6$~ns.

It is important to emphasize that the electrically injected spin
packets are temporally broadened by an amount $\tau_{SB}$. This broadening
becomes particularly important as individual spins start to precess
in $B_z$ at all times during the spin pulse. The retardation of spin
precession results in dephasing of the spin packet and a decrease of
the average spin. The temporal evolution of the packet can be estimated by
\begin{equation}
 M_y(B_z,\Delta t) = \int\limits_0^{\Delta t} dt\ r_S(t)  M_S(\Delta t -
 t),
\end{equation}
\noindent where $r_S(t)=I_p(t)/a$ is the spin injection rate with
the active sample area $a$ and where $M_S$ is given by an
exponentially damped single spin Larmor precession. The integral can
be solved analytically \cite{supplementary} and results in a form as
given by Eq. 1, motivating our original ansatz. Note that the non-precessing
background signal of $\theta_F$ (see Fig. 2(a)) stems from the
discharging of the Schottky capacitance, i.e. $\tau_{SB}=\tau_{bg}$,
while $T_2^*$ is not affected by the integration. The amplitude
$A(B_z)$ in Eq.~1 becomes a function of $\omega_L, T_2^*, \tau_{SB},
\Delta w$ and $r_s$ (see  Eqs. 12/15 in Ref.~\cite{supplementary}).
For simulating $A(B_z)$, we take the results from Fig.~2, i.e.
$T_2^*(B_z)$, $\omega_L$, as well as $\Delta w=2$~ns and vary only
$\tau_{SB}$ as a free parameter. The resulting $A(B_z)$ curves are
plotted in Fig.~2(d) for various $\tau_{SB}$. The experimental data
are remarkably well reproduced for the $\tau_{SB}$ values determined
by TDR ($\tau_{SB}=6$~ns) and by the non-oscillatory background of
$\theta_F$ ($\tau_{SB}=8$~ns). This demonstrates that charging and
discharging of the Schottky capacitance is the main source of the
amplitude drop in our experiment, which limits the observable precession amplitude.

We next explore how this phase coherence, i.e. spin precession, can be
probed by electrical means. Here, we use circularly
($\sigma^+/\sigma^-$) polarized ps laser pump pulses with laser
energy above the GaAs band edge ($E=1.54$~eV) to optically excite
coherent spin packets $\mathbf{S}_{opt}$ in the $n$-GaAs layer
(Fig.~3(d)) near the degenerate region which is formed close to
the Fe/GaAs interface (Fig.~3(a)). The spins generated directly by
this pulse, denoted $\mathbf{S}_{opt}$ in Fig.~3(a), flow into the
GaAs bulk and are oriented perpendicular to the interface. The
optical pulse also generates a photocurrent (PC) which is
spin-polarized due to the replacement of electrons in the
interfacial region by carriers tunneling from the ferromagnet \cite{FI_1,FI_2,FI_3}.  This
spin-polarization, denoted $\mathbf{S}_{pc}$ in Fig.~3(a), is
oriented parallel to the interface. Fig.~3(e) shows the temporal
evolution of the PC as measured by a sampling oscilloscope. As
expected, the PC pulse is temporally broadened by the Schottky
barrier, similar to the injection pulse in TRESI.

The spin-dependent PC $I_{s,pc}$ is measured by a lock-in amplifier
in current detection mode with the laser pump pulses modulated
between $\sigma^+$ and $\sigma^-$ by a photo-elastic modulator at
42~kHz. Furthermore, we reduce the laser repetition time $T_{rep}$
to 12.5~ns.  It is not a priori obvious that $I_{s,pc}$ should
depend on $\mathbf{M}_{Fe}$, since $\mathbf{S}_{opt}$ is
perpendicular to $\mathbf{M}_{Fe}$ and hence $\mathbf{S}_{pc}$.  The
dynamics of the two spin populations, however, are very different.
$\mathbf{S}_{opt}$ is induced by the optical pulse (width $\approx
3$~ps), and it will therefore precess about $B_z$ with a precisely
defined phase for time scales up to $T_2^*,$ which is greater than
50 nanoseconds at low temperatures. The interfacial spin
polarization $\mathbf{S}_{PC}$, however, is generated over the
course of the entire PC pulse (several nanoseconds).

\begin{figure}[tbp]
\includegraphics{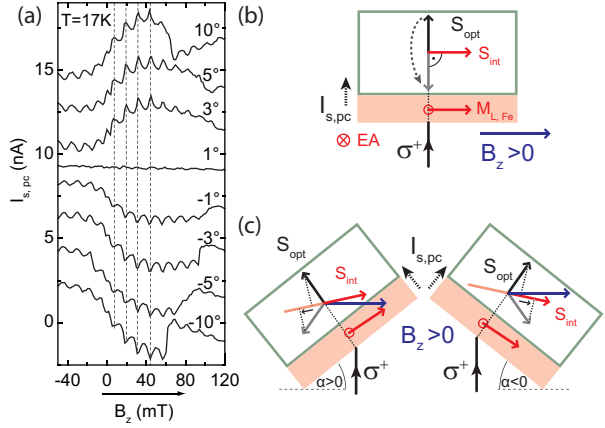}
\caption{ \label{fig3} (color online) (a) Spin PC vs $B_z$ for
different angles  $\alpha$ of the magnetic field relative to the Fe/GaAs interface at 17~K.  (b) Illustration of
sample along $x$-direction under normal incidence ($\alpha=0\deg$)
and (c) for $\alpha>0\deg$ and $\alpha<0\deg$. The spin
polarized PC pulse leads to an interfacial spin
accumulation $\mathbf{S}_{int}$, which
acts as a spin detector for $\mathbf{S}_{opt}$. Spin precession of
$\mathbf{S}_{opt}$ about $B_z$ is observed because of its sensitivity to $\mathbf{S}_{int}$
Note that the sign of the precessing component reverses sign upon reversal of $\alpha$.}
\end{figure}

To probe how these distinct spin populations influence the total PC,
$B_z$ is applied at an angle $\alpha$ relative to the plane of the
Fe/GaAs interface. Fig.~3(f) shows magnetic field loops of
$I_{s,pc}$ with $\alpha = 6\deg$ and the EA of the sample almost
perpendicular to $B_z$. We observe a clear spin-dependence of the PC
signal at all $T$ with abrupt jumps at the coercive fields of the Fe
layer, which occur due to a slight misalignment of the EA
(approximately $4\deg$). At low $T$, peaks in $I_{s,pc}$ are
observed at regularly spaced intervals of 12.87~mT. These are due to
resonant spin amplification \cite{Kikkawa98} of $\mathbf{S}_{opt}$
when the Larmor period is equal the repetition rate of the optical
pulses. In contrast to RSA peaks in TRESI (see Fig.~3(c)), these
peaks survive up to larger $T$ consistent with $T_2^*$ in the GaAs
bulk\cite{Kikkawa98}.

At first glance, RSA oscillations in Fig.~3(f) might be attributed
to the orientation of the precessing spin polarization
$\mathbf{S}_{opt}$ relative to $\mathbf{S}_{pc}$.  The relationship,
however, is more subtle, as can be seen in Fig.~4(a). Remarkably,
when $\alpha = 0$ there is almost no spin-dependent PC, although the
precessing $\mathbf{S}_{opt}$ does have a significant non-zero
projection onto $\mathbf{S}_{pc}$, which is rotating from the easy
toward hard axis as $B_z$ increases. A spin-dependent PC appears
only when the sample is rotated slightly about the EA, as shown for
several angles in Fig.~4(a).  The observed field and angle
dependence indicates that $I_{s,pc}$ is nearly proportional to the
\emph{longitudinal} (non-precessing) component of
$\mathbf{S}_{opt}$, which is oriented along $B_z$. This can only be
due to the fact that over the course of the bulk $T_2^*$-time (the
effective duration of the experiment), the time evolution of
$\mathbf{S}_{pc}$ results in an interfacial spin accumulation
$\mathbf{S}_{int}$ that is nearly parallel to $B_z$, and not to
$\mathbf{M}_{Fe}$.  The spin current that leads to the observed
$I_{s,pc}$ is driven by the gradient between the spin accumulation
in the bulk $S_{opt}$ and the interfacial spin accumulation
$\mathbf{S}_{int}$.

An important question is why the direction of $\mathbf{S}_{int}$
differs from $\mathbf{S}_{pc}$. The primary
reason is dynamic nuclear polarization (DNP), which leads to rapid
precessional dephasing of the transverse component of
$\mathbf{S}_{pc}$ in a hyperfine field that in steady-state points
along $B_z$.  Although this mechanism is well-known\cite{Paget},
a remarkable aspect of the data shown in Fig.~4(a) is that the bulk
spins, as indicated by periodic peaks in $B_z$, continue to precess
at the bare Larmor frequency. This is due to the fact that DNP
occurs only in the interfacial region and not in the bulk, which can
be polarized only by nuclear spin diffusion, which occurs only very
slowly\cite{Stephens2004, Chan2009}.  A second puzzling aspect of
the data is that a precessing component of $\mathbf{S}_{opt}$ is
still observable at lower $T$, implying that $\mathbf{S}_{int}$ is
not exactly parallel to $B_z$ for $\alpha\neq0$ (see Figs.~4(b) and
(c)). There are two possible explanations for this behavior. First,
due to the presence of the Knight field of the spin-polarized
electrons\cite{Chan2009}, the hyperfine field, which dominates the
spin dynamics in the interfacial region, is not exactly parallel to
$B_z$. Second, the fact that the effective duration of the
experiment is several ns means that $\mathbf{S}_{int}$ never reaches
its steady state orientation along $B_z$. A full understanding will
require explicit modeling of the spin dynamics on the ns time-scale,
which is not in the steady-state in contrast to ordinary spin
transport experiments.

In conclusion, we have shown that electrical current pulses can trigger
spin packets that are injected across an Fe/GaAs
Schottky barrier with a well-defined  phase. Charging and discharging of the Schottky barrier
yields a temporal broadening of the spin packets resulting in a
partial dephasing during spin precession. We also have shown that
precession of spin packets generated by optical pulses can be probed electrically due to the spin sensitivity of the resulting
photocurrent. These
results thus demonstrate that electrical injection and detection
of electron spin packets in ferromagnet/semiconductor spin transport
devices are feasible at microwave frequencies.

Supported by HGF and by DFG through SPP 1285. Work at Minnesota was
supported by the Office of Naval Research, in part by the NSF NNIN and MRSEC
programs (DMR 0819885) and NSF DMR 0804244.

\end{document}